\documentclass{article}
\usepackage[utf8]{inputenc}

\title{The CAFAna framework for neutrino analysis}
\author{C. Backhouse -- University College London\\{\tt c.backhouse@ucl.ac.uk}}
\date{25 March 2022}

\usepackage{graphicx}
\usepackage[bookmarks=false]{hyperref}

\usepackage[margin=1in]{geometry}

\begin{document}

\maketitle

The CAFAna framework was initially developed for oscillation analysis with NOvA, and has been adopted for use by the DUNE long-baseline and the SBN short-baseline analyses. In NOvA it is the day-to-day tool for most analysis needs, and is the underlying technology for the published three-flavour oscillation results \cite{nova-3flav}, constraints on sterile oscillations \cite{nova-sterile}, and multiple neutrino cross-section measurements. In DUNE, CAFAna produced the three flavour oscillations sensitivities presented in the TDR \cite{dune-tdr}. Efforts are ongoing to exploit CAFAna for DUNE atmospheric oscillations and for cross-section analyses with the Near Detector. Within SBN, CAFAna is one of three frameworks \cite{valor} considered for oscillation analysis. CAFAna will also be crucial in the planned NOvA-T2K joint fit.

CAFAna takes its name from CAFs -- {\bf C}ommon {\bf A}nalysis {\bf F}iles -- the standard analysis-level file format used throughout NOvA \cite{cafs}. These are conventional ntuples, consisting of a ROOT \cite{root} TTree, in which each record represents a single neutrino candidate, or ``slice''. NOvA has also experimented with an HDF5 \cite{hdf5} representation of the CAFs. The exact details of the format are not crucial, and similar files from DUNE and SBN act as the inputs in those ports.

Fundamentally, CAFAna fills two roles: the filling of histograms from CAFs, and the measurement of oscillation parameters using those histograms. Speed and efficiency are key requirements in each case. The computational demands of oscillation fits are well understood, but minimizing user wait times in exploratory data analysis is also extremely important for productivity.

CAFAna is written in C++ and is driven by the user via the familiar mechanism of ROOT macros. There is also a python interface enabled by pyROOT. The philosophy is to provide a complete set of well-designed building blocks, and allow the user to combine them programmatically  to achieve their specific objectives. The main classes are well documented, users learn the core concepts quickly, and a variety of tutorial workshops have been held in NOvA and DUNE, for which video recordings exist.

The basic CAFAna histogram object is the {\tt Spectrum}. This consists of a series of bin contents, internally manipulated via Eigen \cite{eigen}, but available for display purposes as a ROOT histogram, plus associated exposure information (protons-on-target and/or livetime). Spectra provide a variety of mathematical operations and methods of combination, each of which takes proper account of the exposure information, essentially eliminating proper scaling from the issues of concern to the user. Multi-dimensional spectra are supported natively without
adding significant complexity to the implementation. Internally they are flattened to one dimension for all operations, and reconstituted for display.

In order to allow efficient reading from the input ntuples, CAFAna encourages the user to fill all necessary spectra in a single pass. This allows the CAFAna machinery (the {\tt SpectrumLoader} class) to only visit each record in the file once. The syntax is declaratory; the user specifies the dataset, custom binning, variable ({\tt Var}) and selection ({\tt Cut}) when constructing each {\tt Spectrum}, and then a single operation loops through the input files and fills them all as described. {\tt Cut}s may be combined intuitively, using the C++ logical operators.  {\tt Cut} and {\tt Var} appear to the user as simple functions mapping a neutrino record to a boolean or numeric result respectively. However, CAFAna, through the use of a sub-package called SRProxy, ensures that the full record is never instantiated, and only the fields referenced in the analysis are read from the file. Within each experiment, the large collection of {\tt Cut}s and {\tt Var}s accumulated by analysis users form a valuable commons, which enable new analyses to make a rapid start, building on tried-and-tested estimators and selections.

The SRProxy technology also enables an elegant approach to systematic variations. While many systematics, such as flux and cross-section uncertainties, may be implemented by event reweighting, others, such as various detector effects ({\it e.g.} energy scales) are best represented by modifications of the event record. CAFAna systematic shifts, which are also specified in the definition of each {\tt Spectrum} allow for a weighting, but also a rewriting of each event record. Together, the {\tt SpectrumLoader} and SRProxy machinery ensure that the event record is edited in-place and reverted in an efficient sequence so that each {\tt Spectrum} is ultimately filled from the appropriately modified record.

While the ideal is that the user should be able to produce their spectra in a matter of seconds or minutes, often very large datasets are used or large numbers of spectra are to be produced. Almost every object in CAFAna is persistable, so that once the event reading phase is complete, it need not be repeated. CAFAna is integrated with the Fermilab {\tt jobsub} infrastructure and the SAM data management system, so that a single macro may be run on the grid, each instance accessing a unique subset of the files in a SAM dataset, with no changes required to the interactive macro by the user. CAFAna objects are carefully designed so that the output of such distributed processing is safely combinable ({\it e.g.} a ratio must be stored as its numerator and denominator separately).

Oscillation analyses are based on the {\tt OscillatableSpectrum}, a two-dimensional spectrum of the users' variable ({\it e.g.} reconstructed neutrino energy) versus the true neutrino energy (or true $L/E$) that allows the resulting one-dimensional spectrum from any set of oscillation parameters to be computed by weighting of the true energy bins. These are further combined into more complex objects, such as the {\tt Prediction}, which represents a spectrum to be compared to the data, computable at any combination of oscillation and systematic parameters.

Initially, CAFAna was concerned with an extrapolation-style analysis, where the Far Detector prediction is updated on the basis of Near Detector data/MC comparisons. However, the framework is sufficiently flexible that the DUNE analysis, and NOvA and SBN sterile analyses, instead make use of a joint fit between the two detectors. Similarly, the standard statistical treatment is a frequentist approach with pull terms for systematic uncertainties, using the Feldman-Cousins prescription \cite{fc} to guarantee correct coverage. CAFAna contains tools to provide the predicted spectrum as a function of systematic parameters by interpolating between pre-computed templates, while the DUNE and NOvA sterile analyses mentioned above have contributed various tools for the use of covariance matrices to represent some or all systematic uncertainties, where that trade-off is deemed worthwhile. A larger expansion of the CAFAna scope was a project to additionally enable Bayesian statistical approaches, using the STAN framework for Hamiltonian MCMC \cite{stan}. This required the use of STAN machinery for the analytic computation of derivatives to be retrofitted into various core classes, but the work is now mature and due to be used in upcoming results.

CAFAna is a mature framework with a large user base, so the basic design is essentially fixed. Work continues to further optimize the underlying machinery. For example, the NERSC supercomputer \cite{nersc} used for NOvA and DUNE Feldman-Cousins analysis has unique requirements. There is ongoing work on a variety of extensions such as covariance matrix methods, Bayesian statistics, and adaptations to atmospheric and other analyses that do not map perfectly to the standard long-baseline paradigm. The expansion of CAFAna to DUNE and SBN contexts happened rapidly, and led to some fragmentation of the codebase. An ongoing project is to reunify the various strands of development on top of a common core \cite{cafanacore}, to minimize maintenance burden, ensure that valuable contributions from each experiment benefit all users, and allow for seamless transfer of experience as users move between collaborations.

CAFAna has been an important enabling technology for much important research, including some key recent results in the field. The freedom and flexibility its structure affords users has led it to be the clear choice for day-to-day productivity within NOvA, and the first forays into new frontiers in the last few years have proved the adaptability and general applicability of the approach. CAFAna has benefited from the contributions of uncountable collaborators during its existence so far, and we look forward to continuing in that spirit.

\end{document}